\documentclass[twocolumn]{aastex62}

\graphicspath{{./}{figures/}}
\usepackage[utf8]{inputenc}
\usepackage{graphicx}
\usepackage{dcolumn}
\usepackage{bm}
\usepackage{ulem} 
\usepackage{hyperref}

\usepackage{multirow,enumitem}
\usepackage{amsmath}
\usepackage{color}
\usepackage{xcolor}
\usepackage{multirow}

\newcommand{\LCDM}{\rm{\Lambda}CDM}

\begin{document}


\title{A Model-independent Method to Determine $H_0$ Using Time-Delay Lensing, Quasars and Type Ia Supernovae}

\author{Xiaolei Li}
\affiliation{College of Physics, Hebei Normal University, Shijiazhuang 050024, People's Republic of China}
\author{Ryan E. Keeley}
\affiliation{Department of Physics, University of California Merced, 5200 North Lake Road, Merced, CA 95343, USA}
\author{Arman Shafieloo}
\affiliation{Korea Astronomy and Space Science Institute, Daejeon 34055, Republic of Korea}
\affiliation{University of Science and Technology, Yuseong-gu 217 Gajeong-ro, Daejeon 34113, Republic of Korea}
\author{Kai Liao}
\affiliation{School of Physics and Technology, Wuhan University, Wuhan 430072, People's Republic of China}

\date{\today}
 
\begin{abstract}

Absolute distances from strong lensing can anchor Type Ia Supernovae (SNe Ia) at cosmological distances giving a model-independent inference of the Hubble constant ($H_0$). Future observations could provide strong lensing time-delay distances with source redshifts up to $z\,\simeq\,4$, which are much higher than the maximum redshift of SNe Ia observed so far. In order to make full use of time-delay distances measured at higher redshifts, we use quasars as a complementary cosmic probe to measure cosmological distances at redshifts beyond those of SNe Ia and provide a model-independent method to determine $H_0$.
In this work, we demonstrate a model-independent, joint constraint of SNe Ia, quasars, and time-delay distances {{from strong lensed quasars}}. We first generate mock data sets of SNe Ia, quasar, and time-delay distances based on a fiducial cosmological model. Then, we calibrate the quasar parameters model independently using Gaussian process (GP) regression with mock SNe Ia data. Finally, we determine the value of $H_0$ model-independently using GP regression from mock quasars and time-delay distances from strong lensing systems. As a comparison, we also show the $H_0$ results obtained from mock SNe Ia in combination with time-delay lensing systems whose redshifts overlap with SNe Ia. Our results show that quasars at higher redshifts show great potential to extend the redshift coverage of SNe Ia and thus {{enable}} the full use of strong lens time-delay distance measurements from ongoing cosmic surveys and improve the accuracy of the estimation of $H_0$ from $2.1\%$ to $1.3\%$ {{when the uncertainties of the time-delay distances are $5\%$ of the distance values}}.

\end{abstract}


\section{Introduction}
The simplest flat $\LCDM$ model explains a large range of current observations including cosmic microwave background radiation (CMB), Big Bang nucleosynthesis and baryon acoustic oscillation (BAO) measurements \citep{2009astro2010S.314S,Planck:2013pxb,Planck:2015fie,Planck:2018vyg,2021MNRAS.504.4667A}. However, there are significant tensions between different data sets when $\LCDM$ is used to estimate some key cosmological parameters. 
One of the major issues is the discrepancy between the value of the Hubble constant measured by the {{multiple local-universe probes ~\citep{Riess_2018,Riess_2019,Reid_2019,Riess:2021jrx}} and that inferred by early-universe probes under the assumption of $\LCDM$ cosmology \citep{Planck:2018vyg}. This tension has reached the 4$\sigma$ to 6$\sigma$ level \citep{DiValentino:2021izs}.}

The tension either could be due to unknown systematic errors in the observations or could reveal new physics beyond $\LCDM$. A model-independent method to determine $H_0$ from observations in the redshift gap between {{local-universe probes and early-universe probes is necessary to better assess the $H_0$ tension.}}
Quasars are luminous persistent sources in the Universe which can be observed up to redshifts of $z\simeq7.5$ \citep{2011Natur.474..616M}. The magnifying effect of strong gravitational lensing can be used to observe quasars at even higher redshifts. {{With future surveys, the redshift of SNe Ia from Nancy Grace Roman Space Telescope (ROMAN) SN could reach $z\sim2$ with larger uncertainties \citep{Hounsell:2023xds}. On the other hand, future surveys will provide us with more strong lensing system measurements with higher redshift\citep{2010MNRAS.405.2579O}.} }
Therefore, lensed quasars act as a potential cosmic probe at higher redshifts to {{shrink}} the redshift gap between the farthest observed SN Ia and CMB observations.

Recently, a feasible method to determine $H_0$ independent of the cosmological model that used strong lensed quasars and Type Ia supernovae (SNe Ia) with Gaussian Process (GP) regression has been presented in \cite{Liao:2019qoc,Liao:2020zko}.
Strong gravitational lensing of a variable source measures the time-delay distance $D_{\rm{\Delta t}}$ of the system and measuring the stellar velocity dispersion of the lens also yields a constraint on the angular diameter distance to the lens $D_{\rm{d}}$.
One can anchor SNe Ia with these absolute distances and obtain an excellent constraint on the shape of the distance-redshift relation
~\citep{2019PhRvL.123w1101C}. In \cite{Liao:2019qoc,Liao:2020zko}, the authors applied GP regression to SNe Ia data to get a model-independent relative distance-redshift relation and anchored the distance-redshift relation with $D_{\rm{\Delta t}}$ and $D_{\rm{d}}$ from strong gravitational lensing to give the constraints on $H_0$.
 
However, observations of strong lens systems summarized in \cite{DES:2022dvw} show that the redshift of lensed quasars could reach as high as $z\simeq3.8$, which is far beyond the highest redshift of observed SNe Ia so far. Therefore, looking for observations at higher redshifts is necessary to make full use of time-delay lensing systems.
Recently \cite{Du:2023zsz} used gamma-ray burst (GRB) distances and H0LiCOW lenses with redshifts $<1.8$ to infer $H_0$.

Moreover, the linear relation between the $\log$ of the UV and X-ray luminosities allows quasars to be potentially used as standard candles at higher redshifts if well calibrated \citep{Risaliti:2015zla,Lusso:2017hgz,Risaliti:2018reu,Lusso:2020pdb,Khadka:2020tlm,Khadka:2020vlh,Li:2021onq}. Thus, the combination of time-delay observations in strong lensed quasars and the linear relation between the $\log$ of the ultraviolet (UV) and X-ray luminosities of quasars can help us to determine $H_0$ model-independently. 


In our work, we use GP regression to reconstruct the expansion history of the Universe model-independently. 
We first generate SNe Ia, quasar, and strong lens data set based on a fiducial cosmological model. Then, we calibrate the mock quasar data set by using GP regression to {{model-independently}} reconstruct the expansion history of the Universe from the mock SNe Ia data set following the previous work by \citep{Li:2021onq}. Using the calibrated quasar data set, we further reconstruct the expansion history up to redshift of $z\simeq7.5$ with GP. Then following \cite{Liao:2019qoc,Liao:2020zko} we determine $H_0$ cosmological-model-independently using simulated strong lensed quasars with source redshifts up to 4 and calibrated {{unlensed}} quasars. 

This paper is organized as follows: in Section~\ref{sec:data}, we describe the data sets we used in detail. The quasar calibration with GP regression from the latest SNe Ia observations, as well as the determination of $H_0$ from strong lens systems and calibrated quasars are shown in Section~\ref{sec:method}. We discuss our conclusions in Section~\ref{sec:con}.

\section{Data} \label{sec:data}
{{Since there is a lack of necessary time-delay measurements for strong lensing systems, we are going to simulate the time-delay measurements based on a fiducial cosmological model. Moreover, to make the results convictive, we use simulated SNe Ia data as well as simulated quasar samples instead of the real data from observations.}}

In this section, we briefly describe the method of generating the mock data based on a fiducial cosmological model. Throughout our work, a flat-$\LCDM$ model with $\Omega_{\rm{m}}= 0.3$ and $H_0 = 70 \, {\rm{km\,s^{-1}\,Mpc^{-1}}}$ is used as the fiducial cosmological model. We should emphasize here that following our previous works, we could have selected any cosmological model (as the fiducial model) for our analysis since we are performing a model-independent analysis for reconstructing the expansion history. We have chosen the standard flat-$\LCDM$ model as our fiducial model since the focus of this paper is on the high precision determination of $H_0$ using high redshift quasars.  

\subsection{Type Ia Supernovae}
SNe Ia, which helped discover cosmic acceleration, are powerful standard candles that enable precise measurements of the expansion of the Universe.
{{A most recent Pantheon+ sample was reported in \citep{Scolnic:2021amr} which consists of 1701 light curves of 1550 distinct SNe Ia ranging in redshift from z=0.001 to 2.26. This larger SNe Ia sample is a significant increase compared to the original Pantheon sample, especially at lower redshifts.}}

In this work, we generate a mock SNe Ia data set based on the Pantheon+ sample assuming a fiducial cosmological model.
First, we obtain the luminosity distances of SNe Ia with
\begin{equation}\label{eq:DL}
    D_{\rm{L}}^{\rm{fid}}(z)\,=\,\frac{c(1+z)}{H_0} \int_0^z \frac{dz}{\sqrt{\Omega_m(1+z)^3+(1-\Omega_m)}}
\end{equation}
and then the distance modulus can be calculated with
\begin{equation}\label{eq:DM}
    \mu^{\rm{fid}}\,=\,5{\log(\frac{D_{\rm{L}}^{\rm{fid}}}{{1 {\rm{Mpc}}}})}+25.
\end{equation}
The mock SNe Ia data, $\mu^{\rm{mock}}$, are then generated from $\mu^{\rm{fid}}$ by adding noise as a random variable with a mean of zero and a variance characterized by the Pantheon+ covariance matrix. 
We use this mock SNe Ia data set along with the mock quasar data set to simultaneously calibrate the mock quasar data set and reconstruct the Universe's expansion history.

\subsection{Quasar sample}
Quasars act as standard candles based on the log-linear relation between the UVt and the X-ray luminosities $\log(L_{\rm{X}})=\gamma\log(L_{\rm{UV}})+\beta_1$. This allows quasars to work as cosmic probes at higher redshifts to {{shrink}} the redshift gap between SNe Ia and the CMB if well calibrated since {{they}} can be observed up to the redshifts of $z\simeq 7.5$. So far, the largest quasar sample with both X-ray and UV observations consists of $\sim 12,000$ objects. However, after applying several filtering steps to reduced the systematic effects, 2421 quasars with spectroscopic redshifts and X-ray observations from either Chandra or XMM–Newton in the redshift range of $0.009<z<7.54$ were left in the final cleaned sample \citep{Lusso:2020pdb}.

In this work, we generate a mock quasar data set based on the quasar catalog described above assuming a fiducial cosmological model. First, we take the values of $\log (F_{\rm{UV}})^{\rm{fid}}$ from the actual measurements. Then, we calculate $\log (F_{\rm{X}})^{\rm{fid}}$ using
\begin{equation} \label{eq:logFx1}
    {\log (F_{\rm{X}})^{\rm{fid}}}\,=\,\gamma \log (F_{\rm{UV}})^{\rm{fid}} + (2\gamma-2){\rm{log }}(D_{\rm{L}}^{\rm{fid}}) + \beta_2
\end{equation}
where $\beta_2 = \gamma \log (4\pi)-\log (4\pi)+\beta_1$, $F_{\rm{UV}}$ and $F_{\rm{X}}$ are the fluxes measured at fixed rest-frame wavelengths, and $D_{\rm{L}}^{\rm{fid}}$ is the luminosity distance relation of the fiducial cosmology. 
$\gamma$ and $\beta_2$ are quasar parameters that need to be calibrated. These calibration parameters are degenerate with the cosmological parameters, or model-independent distances we want to fit or reconstruct. Since $\beta_2$ is degenerate with $H_0$, quasars can only measure relative distances, just like SNe Ia. Thus, we absorb $H_0$ into the parameter $\beta = \beta_2 - (2\gamma -2)\log (H_0)$. This is to absorb multiple degenerate parameters which characterize the relative anchoring between the data and the expansion history into one parameter. Here, we use fiducial values for $\gamma = 0.6430$
and $\beta = 7.88$, which are the best-fit values from \cite{Li:2021onq}.  
The final sample of fluxes ($\log (F_{\rm{UV}})^{\rm{mock}}$, $\log (F_{\rm{X}})^{\rm{mock}}$) are calculated from the fiducial values ($\log (F_{\rm{UV}})^{\rm{fid}}$, $\log (F_{\rm{X}})^{\rm{fid}}$) by adding Gaussian random noise with a standard deviation ($\sigma_{\log (F_{\rm{UV}})}$, $\sigma_{\log (F_{\rm{X}})}$) from the actual data set.

\subsection{Strong lens time-delay distance data}

A typical strongly lensed system, as used for time-delay cosmography, consists of a source quasar
at cosmological distances, which is lensed by a foreground elliptical galaxy, 
and forms multiple images of the quasar and the arcs of the host galaxy. 
With years of observations of the light curves, one can measure the time delay between any two images, which, following the Fermat principle, arises from the different geometries and Shapiro time delays along the {{multiple}} paths. 
The time delay thus depends on both the geometry of the Universe and the gravitational field of the lens galaxy.
The time delays $\Delta t$ can be used to measure a time-delay distance $D_{\Delta t}$ following
\begin{equation}
    \Delta t\,=\,D_{\Delta t} \Delta \phi (\xi_{\rm{lens}})
\end{equation}
where $\Delta \phi $ is the Fermat potential difference between the two images which is a function of lens mass profile parameters $\xi_{\rm{lens}}$, determined by high-resolution imaging of the host arcs.
$D_{\Delta t}$ is the time-delay distance
\begin{equation}\label{eq:TDD}
      D_{\Delta t}\,=\,(1+z_{\rm{d}}) \frac{D_{\rm{d}}D_{\rm{s}}}{D_{\rm{ds}}}
\end{equation}
which is a combination of three angular diameter distances $D_{\rm{d}}$, $D_{\rm{s}}$, and $D_{\rm{ds}}$ where the subscripts d and s denotes the deflector (lens) and the source, respectively. Time-delay distance measurements can be used as a one-rung distance ladder and are independent of the Cepheid distance ladder and early Universe physics. The angular diameter distance to the deflector lens itself, $D_{\rm d}$, can be obtained independently of the time-delay distance of the strong lens system. These $D_{\rm d}$ measurements provide additional constraints on the expansion history beyond the $D_{\Delta t}$ measurements. 

With the increasing number of the wide-field imaging surveys, e.g., {{Vera C. Rubin Observatory's Legacy Survey of Space and Time (LSST), Euclid, and ROMAN}}, the number of the known strong lens systems is growing rapidly \citep{2015arXiv150303757S,2019arXiv190205569A,2010MNRAS.405.2579O,2015ApJ...811...20C}. New images of billions of galaxies are expected to be observed, of which {{$\sim 100,000$}} are strong lens systems \citep{Collett:2015roa}.

In this work, we simulate angular diameter distances and time-delay distances for each strong lens system following {{the}} method described below:
\begin{enumerate}
\item  In the fiducial cosmological model, the angular diameter distance can be calculated via
\begin{equation}
    D^{\rm{fid}}_{\rm{A}}(z)\,= D_{\rm L}^{\rm{fid}} / (1+z)^2
\end{equation}
with the redshift to the deflector $z_{\rm{d}}$ 
and the redshift to the source $z_{\rm{s}}$, we can calculate $D_{\rm{d}}$ and $D_{\rm{s}}$, respectively. Random noise following a normal distribution is added to the angular diameter distance. We analyze two cases, one where the noise being added is at the level of 5\% and another at the level of 10\%. {{The value depends on specific systems, observational conditions, algorithm and most importantly systematical errors. We adopted 5\% for the best case and 10\% for the worse case.}}
 
In a spatially flat universe, the distance between the lens and the source ($D_{\rm{ds}}$) is calculated via
\begin{equation}
    D_{\rm{ds}} \,=\,D_{\rm{s}}-\frac{1+z_{\rm{d}}}{1+z_{\rm{s}}}D_{\rm{d}}.
\end{equation}

\item Then, the time-delay distance can be obtained with Equation~(\ref{eq:TDD}).

\item We consider two cases, where the uncertainties on the simulated distances (both $D_{\rm d}$ and $D_{\Delta t}$) are taken to be $5\%$ and $10\%$.

\end{enumerate}

In this work, we use nine mock strongly lensed quasars based on those in \cite{Ertl:2022rqx} which is a subset of 30 quadruply imaged quasars in \cite{DES:2022dvw}. {{The sample will probably be analyzed by the TDCOSMO team in the next stage \citep{Treu:2022aqp}.}} The redshifts of strong lens systems are summarized in Table~\ref{tab:GL_z} which is part of Table A.3 from \cite{Ertl:2022rqx}. These systems have higher source redshifts than H0LiCOW and might be well analyzed by the TDCOSMO team in the near future. The boldface shows the lensing systems whose redshifts in the redshift range of SNe Ia ($z<2.261$). As can be seen from Table~\ref{tab:GL_z}, only three strong lens systems are left in the redshift coverage of SNe Ia.

Following the simulation method described above, we obtain the simulated angular diameter distance to the lens and the time-delay distance, which are shown in the last two columns of Table~\ref{tab:GL_z}. Throughout our work, we consider $5\%$ and $10\%$ of distance values as uncertainties for the mock time-delay distance $D_{\rm{\Delta t}}$ and angular diameter distance to the deflector $D_{\rm{d}}$.
\begin{table}[]
    \caption{The deflector and source redshifts and the simulated angular diameter distance and time-delay distance for our strong lens systems. We denote lensing systems whose redshifts are overlap with the redshifts of SNe Ia ($z<2.261$) in boldface.}
    \centering
    \begin{tabular}{c| c c| c c}
    \hline
    System    &   $z_{\rm{d}}$ & $z_{\rm{s}}$  & $D_{\rm{d}}$ &$D_{\Delta t}$ \\
    \hline
    DES J0029-3814 & 0.863   & 2.821                                  &$1523.3$ & $ 6315.0$\\
    DES J0214-2105 & 0.22    & 3.229                                  &$801.4 $ & $ 988.6 $\\
    DES J0420-4037 & 0.358   & 2.4                                    &$1030.8$ & $ 1907.7$\\
    PS J0659+1629  & 0.766   & 3.083                                  &$1420.3$ & $ 4976.3$\\
    2M1134-2103    & 0.5     & 2.77                                   &$1273.9$ & $ 2732.5$\\
    {\textbf{J1537-3010}}     & {\textbf{0.592}}   & {\textbf{1.721}} &$1326.7$ & $ 4199.4$\\
    {\textbf{PS J1606-2333}}  & {\textbf{0.5}}     & {\textbf{1.69}}  &$1234.7$ & $ 3562.1$\\
    PS J1721+8842  & 0.184   & 2.37                                   &$628.9 $ & $ 838.8 $\\
    {\textbf{DES J2100-4452}} & {\textbf{0.203}}   & {\textbf{0.92}}  &$780.0 $ & $ 935.5 $\\
    \hline
    
    \end{tabular}

    \label{tab:GL_z}
\end{table}

\begin{figure*}
\centering
\includegraphics[width=0.90\textwidth]{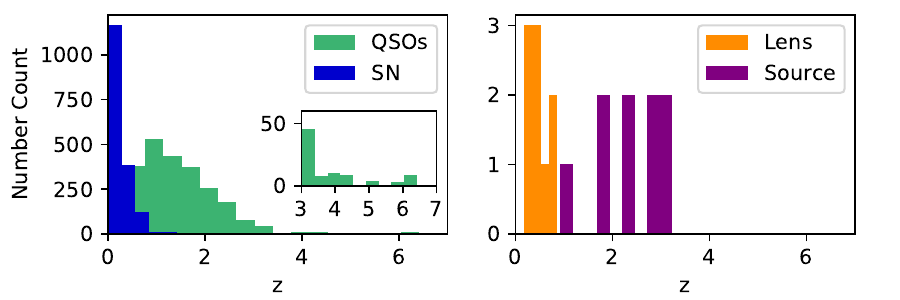}
\caption{{{The left plot shows the redshift distribution of the Quasar sample from \cite{Lusso:2020pdb} and SNe Ia from Pantheon+ sample \citep{Scolnic:2021amr} while the right plot shows the redshift distribution of the sources and deflector for the strong lens time-delay data set from \cite{Ertl:2022rqx}. The redshift distribution of quasars at higher redshift are also displayed in the inner plot to make it more clear.}}}
\label{fig:redshift_distri}
\end{figure*}

We show the redshift distribution for the data set used in our analysis in Figure~\ref{fig:redshift_distri}.

\begin{figure*}
\centering
\includegraphics[width=0.75\textwidth]{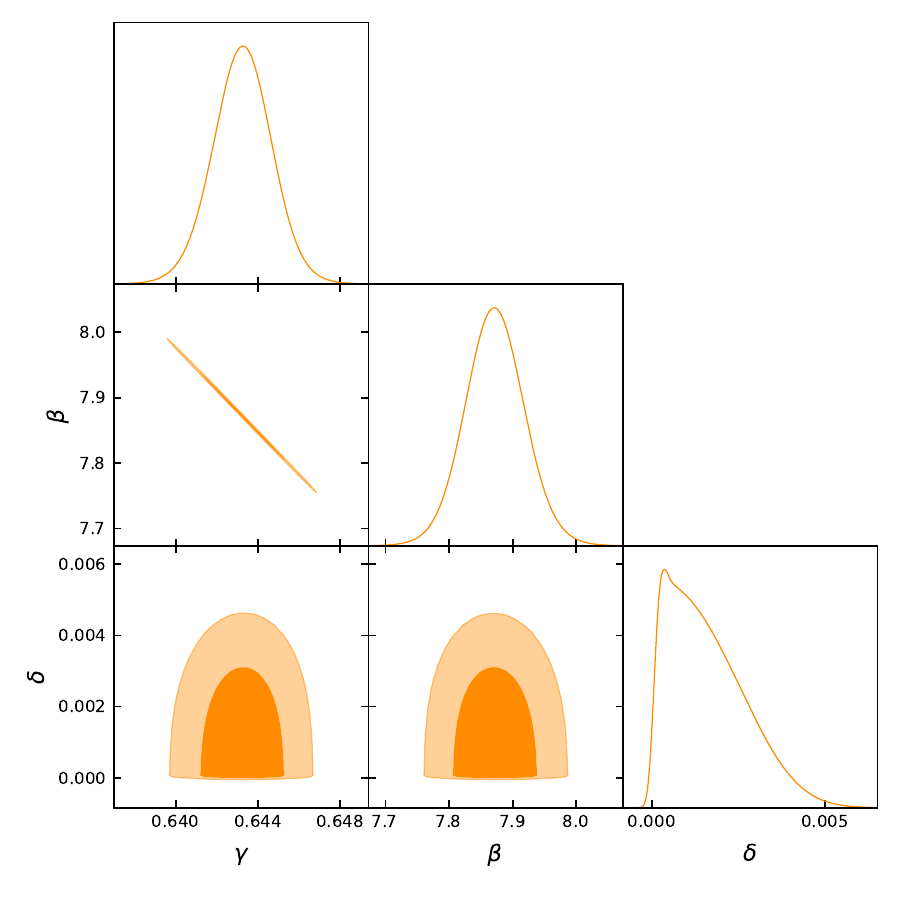}
\caption{Model-independent calibration results for the quasar parameters, 
{{where $\gamma$ is the slope of the log-linear relation between the UV and X-ray luminosity of quasars and $\beta$ relates to the intercept of the relation. While $\delta$ is the intrinsic scatter of quasars.}}
GP reconstructions of $D_LH_0$ based on the mock SNe Ia data.
The contours represent the 1$\sigma$ and 2$\sigma$ uncertainties for $\gamma , \beta,$ and $\delta$.}
\label{fig:quasar_calibration_res}
\end{figure*}

\begin{figure}
\centering
\includegraphics[width=0.45\textwidth]{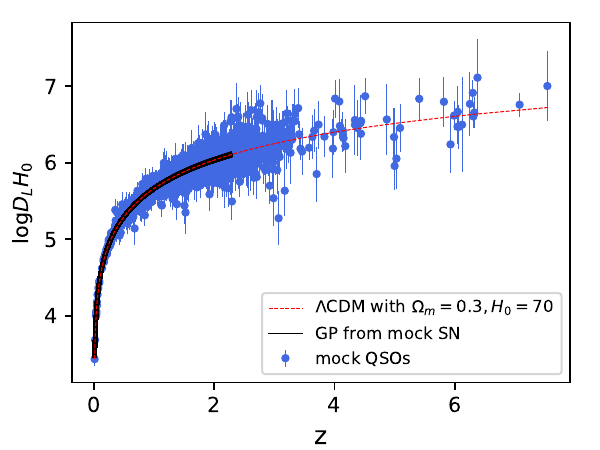}
\caption{$\log (D_{\rm{L}}H_{\rm{0}})$-redshift relation for the mock quasars. The errorbars of $\log (D_{\rm{L}}H_{\rm{0}})$ are obtained through error propagation and the black solid lines show $\log (D_{\rm{L}}H_0)$ obtained from mock SNe Ia data and the dashed red line denotes flat $\LCDM$ model with $H_0 = 70 {\rm{km\,s^{-1}\,Mpc^{-1}}}$ and $\Omega_m = 0.3$ as comparison.}
\label{fig:calibratedquasars}
\end{figure}

\begin{figure}
\includegraphics[width=0.45\textwidth]{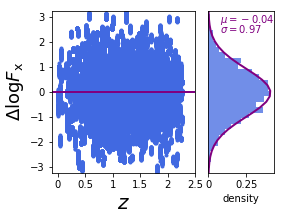}
\caption{Residuals of the mock $\log  (F_{\rm{X}})$ values with respect to the predicted $\log  (F_{\rm{X}})$ values derived from the GP reconstructions of the mock SNe Ia compilation, normalized to the calibrated errors (observational and intrinsic). The right plot shows the histogram for $\Delta \log  (F_{\rm{X}})$ and the purple line shows the best Gaussian fit with $\mu = -0.04$ and $\sigma=0.97$. }
\label{fig:residual}
\end{figure}

\begin{figure*}
\centering
\includegraphics[width=0.45\textwidth]{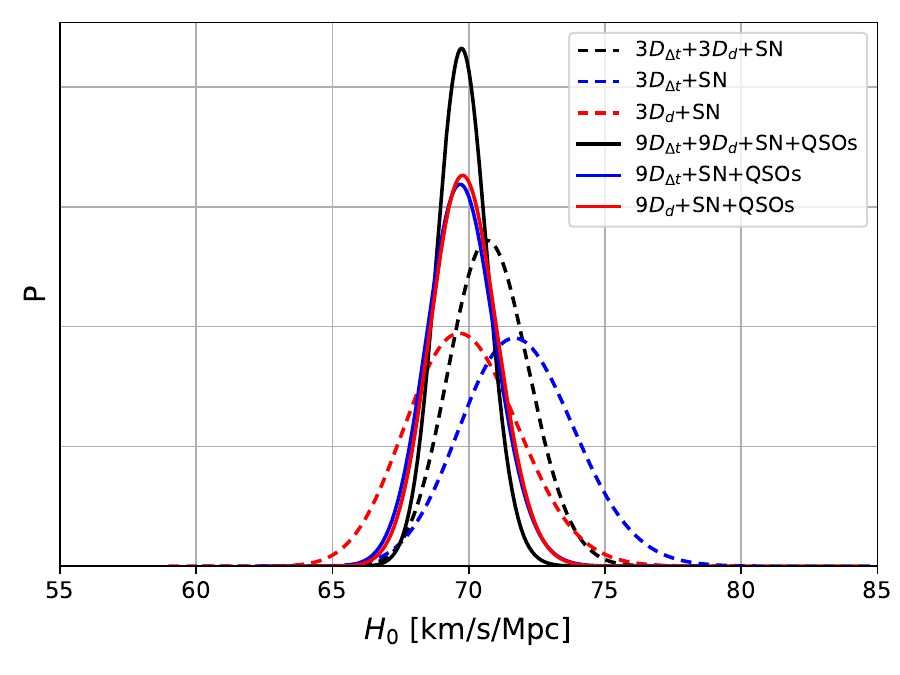}
\includegraphics[width=0.45\textwidth]{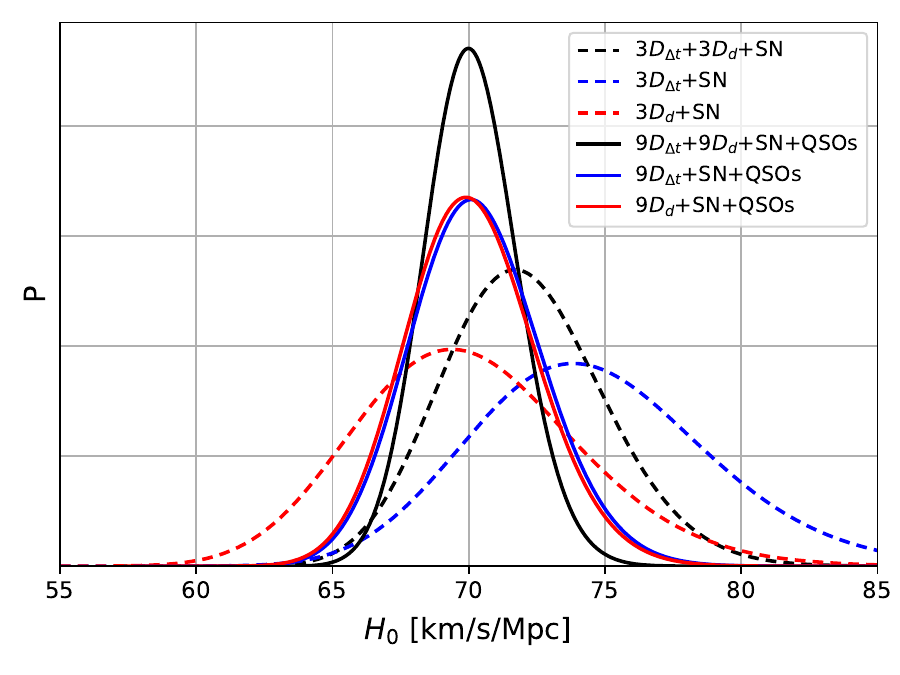}
\caption{The posterior on $H_0$ obtained with simulated time-delay lensing distance ($D_{\Delta t}$) as well as lensing distance ($D_d$). The left plot denotes the results with $5\%$ uncertainties of the simulated distance values and the right plot denotes the results with $10\%$ uncertainties when doing simulation. }
\label{fig:H0_posterior}
\end{figure*}

\section{Methods and Results} \label{sec:method}

\subsection{Quasar calibration}
In this subsection, we briefly describe the method we used to simultaneously calibrate the quasar sample and reconstruct the expansion history using the SNe Ia data set and GP regression.

In order to calibrate the quasar parameters in a model-independent way, we use cosmological distances from another cosmic probe -- SNe Ia. Since the absolute brightness of SNe Ia is degenerate with $H_0$, only the dimensionless, unanchored luminosity distances ($D_LH_0$) can be constrained. We rewrite Eq.~(\ref{eq:logFx1}) as
\begin{equation} \label{eq:logFx}
    \log (F_{\rm{X}})\,=\, \gamma \log (F_{\rm{UV}}) + (2\gamma-2){\rm{log }}(D_{\rm{L}}H_0) + \beta
\end{equation}
where $\beta=\beta_2-(2\gamma-2)\log (H_0)$. $D_{\rm{L}}H_0$ is reconstructed from the mock SNe Ia data set using GP regression. 
GP regression works by generating a random set of cosmological functions whose statics are characterized by a covariance function. We follow some previous works and use a squared-exponential kernel for the covariance function ~\citep{Rasmussen:2006,Holsclaw:2010nb,Holsclaw:2010sk,2011PhRvD..84h3501H,Shafieloo2012Gaussian,2013PhRvD..87b3520S,2018PhRvD..97l3501J,2023JCAP...02..014H} 
\begin{equation}\label{eq:kernel}
   <\varphi(s_i)\varphi(s_j)>\,=\,\sigma_f^2 \exp\left({-\frac{|s_i-s_j|^2}{2\ell^2}}\right)
\end{equation}
where $s_i=\ln (1+z_i)/\ln (1+z_{\rm{max}})$ and $z_{\rm{max}}=2.261$ is the maximum redshift of the SNe Ia sample.  $\sigma_f$ and $\ell$ are two hyperparameters that are marginalized over.
$\varphi$ is just a random function drawn from the distribution defined by the covariance function of Equation~(\ref{eq:kernel}) and we take this function as $\varphi(z)=\ln \left(H^{\rm{mf}}(z)/H(z)  \right)$, i.e. the logarithm of the ratio between the reconstructed expansion history, $H(z)$, and a mean function, $H^{\rm{mf}}(z)$, which we choose to be the best-fit $\LCDM$ model from the Pantheon+ data set. 
{{The mean function plays an important role in GP regression and the final reconstruction results are not quite independent of the mean function, however, it has a modest effect on the final reconstruction results because the values of hyperparameters help to trace the deviations from the mean function \citep{Shafieloo2012Gaussian,2013PhRvD..87b3520S,2017JCAP...09..031A}. 
Moreover, the true model should be very close to the flat $\LCDM$ model so it is reasonable to choose the best-fit flat $\LCDM$ model from Pantheon+ as a mean function.}} This choice allows us to perform a test of whether the data need some additional flexibility to fit the data beyond the $\LCDM$ model~\citep{Keeley:2020aym}. 

For the details of the reconstruction with GP, we refer the readers to \citep{Rasmussen:2006,Holsclaw:2010nb,Holsclaw:2010sk,2011PhRvD..84h3501H,Shafieloo2012Gaussian,2013PhRvD..87b3520S,2017JCAP...09..031A,Keeley:2020aym,Li:2021onq,2023JCAP...02..014H}.  

With the measurements of $F_{\rm{UV}}$ {{from the quasar sample}} and $D_{\rm{L}}H_0$ from SNe Ia, we obtain ${\rm{log }}(F_{\rm{X}})^{\rm{SN}}$ following Equation~(\ref{eq:logFx}). This allows us to compare the quasar data set and the SNe Ia data set with
\begin{equation}\label{eq:chi2}
\begin{aligned}   
    \ln{\mathcal{L}}\,=\,-\frac{1}{2}\sum_i  \Bigg[ & \frac{\left(\log (F_{\rm{X}}(\gamma,\beta))^{\rm{SN}}_i- \log (F_{\rm{X}})_i^{\rm{QSO}}\right)^2} {s_i^2}  \\ 
    & +{\rm{ln}} (s_i^2)\Bigg]
\end{aligned}
\end{equation}

where $s_i^2\,=\,\sigma_{\log (F_{\rm{X}})}^2+\gamma^2 \sigma_{\log (F_{\rm{UV}})}^2+\delta^2$. 
The intrinsic dispersion $\delta$ of the $L_{\rm{X}}-L_{\rm{UV}}$ relation models various unknown physical properties that scatter the observed $\log(L_{\rm X})$-$\log(L_{\rm UV})$ trend by more than the measurement uncertainty \citep{Risaliti:2018reu,Lusso:2020pdb}.

We then calculate the posterior distribution of the quasar parameters: the slope $\gamma$, the intercept $\beta$ and the intrinsic dispersion parameter $\delta$. We should note that the Hubble constant $H_0$ is absorbed into the parameter $\beta$. This is to absorb multiple degenerate parameters that characterize the relative anchoring between the data and the expansion history into one parameter.
Based on the method described above, we use a Python package named \textit{emcee} \citep{foreman2013emcee} to do the Markov Chain Monte Carlo analysis and flat priors are used for each parameter. 

With the calibration method described above, we obtained the best fit of quasar parameters, $\gamma = 0.643\pm 0.002$,
$\beta = 7.872\pm 0.047$ and
$\delta = 0.0016^{+0.0006}_{-0.0015}$ and the contours are shown in Figure ~\ref{fig:quasar_calibration_res}.

To make sure that our calibrated results give reasonable information about cosmology, we calculate $\log(D_LH_0)$ versus $z$ relation from the quasar fluxes with the calibrated quasar parameters through
\begin{equation} \label{eq:logDLH0}
{\rm{log}}(D_{\rm{L}}H_0)\,=\,\frac{\log (F_{\rm{X}})-\gamma \log (F_{\rm{UV}})-\beta}{(2\gamma-2)}.
\end{equation} 
The results are shown in Figure~\ref{fig:calibratedquasars}. The blue points represent the unanchored distances from the calibrated quasar sample, which we use to anchor time-delay distances of strong lensing in later work.  In Figure~\ref{fig:calibratedquasars} we also show the $\log(D_LH_0)$ obtained from the posterior of SNe Ia calculated with GP.

Moreover, we check the consistency between the calibrated quasar sample and the unanchored luminosity distance from SNe Ia by estimating the normalized residual of $\log (F_{\rm{X}})^{\rm{SN}}$ which is calculated via
\begin{equation}\label{eq:residual}
    \Delta \log (F_{\rm{X}})\,=\,\frac{\log (F_{\rm{X}})^{\rm{SN}}-\log (F_{\rm{X}})^{\rm{{QSO}}}}{\sqrt{\sigma_{\log (F_{\rm{X}})}^2+\gamma^2 \sigma_{\log (F_{\rm{UV}})}^2+\delta^2}}.
\end{equation}
where $\log (F_{\rm{X}})^{\rm{SN}}$ is {{obtained}} with Equation~(\ref{eq:logFx}) and $\log (F_{\rm{X}})^{\rm{{QSO}}}$ is the quasar measurements.
The results for the residual are shown in Figure~\ref{fig:residual}. From Figure~\ref{fig:residual} we can see that the distribution of the normalized residual is a Gaussian distribution, which indicates that the $\log(F_{\rm{X}})$ data from quasar measurements is consistent with that derived from SNe Ia using calibrated quasar parameters.

Finding internal consistency between the calibrated quasar and SNe Ia data sets would show that quasars can be used as standard candles at higher redshift and {{are therefore}} powerful probes of cosmology. {{We also need to emphasize here that there are deviations from the standard $\LCDM$ model for quasars at higher redshift as standard candles. It is not clear so far that whether the deviations are due to new physics beyond the $\LCDM$ model or the evolution of calibration parametrizations. Hopefully, future surveys will provide us more copious and precise data for quasars which could help us solve this puzzle. }
}

\subsection{$H_0$ determination}

In order to determine $H_0$ with our technique, we have to take the unanchored reconstructions of the expansion history from SNe Ia and quasars, which only measure relative distances, and anchor them with the strong lens data set, which does measure absolute distances.
We first generate 1000 posterior samples of the $H_0$-independent quantity $D_LH_0$ from quasars
with GP regression mentioned above and convert these unanchored luminosity distances to unanchored angular diameter distances $D^{\rm{A}}H_0$. Then we evaluate the values of each of the 1000 $D^{\rm{A}}H_0$ curves at the lens and the source redshifts of the simulated strong lens systems to calculate 1000 values of $H_0D_{\Delta t}$ using
    \begin{equation}
       H_0D_{\Delta t}\,=\,(1+z_{\rm{d}})\frac{(H_0D_{\rm{d}})(H_0D_{\rm{s}})}{(H_0D_{\rm{ds}})}
    \end{equation}
where $D_{\Delta t}$ is the time-delay distance. Comparing 1000 $H_0D_{\Delta t}$ and $H_0D_d$ curves with simulated $D_{\Delta t}$ and$D_{d}$ at the lens and source redshifts of the lensing systems, we calculate the likelihood with
    \begin{equation}
        (\ln{\mathcal{L}})_{D_{\Delta t}}\,=\,-\frac{1}{2}\sum \frac{\frac{H_0D_{\Delta t}}{H_0}(H_0;z_{\rm{d}},z_{\rm{s}})-D_{\Delta t}^{\rm{sim}}(z_{\rm{d}},z_{\rm{s}})}{\sigma_{D_{\Delta t}}(z_{\rm{d}},z_{\rm{s}})}
    \end{equation}
    and 
        \begin{equation}
        (\ln{\mathcal{L}})_{D_{\rm{d}}}\,=\,-\frac{1}{2}\sum \frac{\frac{H_0D_{\rm{d}}}{H_0}(H_0;z_{\rm{d}})-D_{\rm{d}}^{\rm{sim}}(z_{\rm{d}})}{\sigma_{D_{\rm{d}}}(z_{\rm{d}})}
    \end{equation}
 In the end, we marginalize over the realizations to form the posterior distribution of $H_0$.

\begin{table*}[]
    \caption{The best-fit values for $H_0$ and the corresponding 1$\sigma$ uncertainties as well as the precision of the estimation when the precision of the measurement of the distances in the strong lens data set is $5\%$ and $10\%$. For comparison, we include the case where we are limited to using the strong lens systems with source redshifts within the SNe Ia redshift range.}
    \centering
    \begin{tabular}{c|c|c| c | c}
    \hline
    \multirow{2}*{data} & \multicolumn{2}{c|}{$5\%$ Uncertainties} &  \multicolumn{2}{c}{$10\%$ Uncertainties} \\ \cline{2-5}
    & Best-fit Values & Precision & Best-fit Values & Precision \\
    \hline
  $9D_{\rm{d}}+9D_{\Delta t}$+SN+QSO & $69.8\pm 0.9$   &  $1.3\% $  & $70.1\pm 1.7$   &  $2.4\% $   \\
  $3D_{\rm{d}}+3D_{\Delta t}$+SN  & $70.8\pm 1.5$   &  $2.1\% $  & $72.0^{+2.7}_{-3.2}$  & $4.1\% $    \\
  \hline
  $9D_{\Delta t}$+SN+QSO             & $69.7\pm 1.3$   & $1.8\% $  &  $70.3^{+2.2}_{-2.6}$ & $3.4\% $ \\
  $3D_{\Delta t}$+SN              &  $71.9^{+2.0}_{-2.2}$  & $3.0\% $  &  $74.6^{+3.8}_{-4.8}$ &  $5.8\% $ \\
  \hline
  $9D_{\rm{d}}$+SN+QSO               &  $69.8\pm 1.2$  & $1.7\% $  &  $70.1^{+2.2}_{-2.5}$ & $3.4\% $  \\
  $3D_{\rm{d}}$+SN                &  $69.8^{+1.9}_{-2.1}$   & $3.0\% $  & $70.1^{+3.6}_{-4.5}$  & $5.8\% $  \\
    \hline
    \end{tabular}
    \label{tab:H0_values}
\end{table*}

The posterior on $H_0$ in a flat $\LCDM$ model obtained with nine simulated strong lens systems in combination with quasars and SNe Ia
are shown in solid lines in Figure~\ref{fig:H0_posterior}. The left plot shows the $H_0$ estimation results taking $5\%$ of the distance values as uncertainties while the right plot shows the $H_0$ estimation results taking $10\%$ of the distance values as uncertainties. 
In Table~\ref{tab:H0_values} we summarize the numerical results. First, we consider the combination of $D_{\Delta t}$ and $D_{\rm{d}}$ from the nine simulated lensing systems. The method described above yields $H_0 = 69.8\pm 0.9 \, {\rm{km\,s^{-1}\,Mpc^{-1}}}$ when taking $5\%$ of the distance values as uncertainties and $H_0 = 70.1\pm 1.7 \, {\rm{km\,s^{-1}\,Mpc^{-1}}}$ when taking $10\%$ of the distance values as uncertainties. 
In addition, we give the results from $D_{\rm{d}}$ and $D_{\Delta t}$ separately to quantify the contribution of $D_{\rm{d}}$ in Figure~\ref{fig:H0_posterior}. The constraints when the angular diameter distances to the lens deflector {{($D_{d}$)}} are considered separately from the time-delay distances {{($D_{\Delta t}$)}} are largely equivalent. {{We can also see from Figure~\ref{fig:H0_posterior} that the constraining power of the  combination of $D_{\rm{d}}$ and $D_{\Delta t}$ will improve a lot compared to the results from $D_{\rm{d}}$ and $D_{\Delta t}$ separately. }}

As a comparison, we also use SNe Ia as standard candles following the same $H_0$ determination method described above to determine $H_0$ model-independently. However, only three strong lens systems are left in the redshift range of SNe Ia.  The constraints are shown in dashed lines in Figure~\ref{fig:H0_posterior} and the best-fit values together with the 1$\sigma$ uncertainties are summarized in Table~\ref{tab:H0_values}. We obtain $H_0 = 70.8\pm 1.5 \, {\rm{km\,s^{-1}\,Mpc^{-1}}}$ taking $5\%$ of the distance values as uncertainties and $H_0 = 72.0^{+2.7}_{-3.2} \, {\rm{km\,s^{-1}\,Mpc^{-1}}}$ taking $10\%$ of the distance values as uncertainties with three time-delay lensing systems in combination with SNe Ia. We see that by using quasars and not just SNe Ia as standard candles, more time-delay lensing systems can be included because quasars are measured out to higher redshift, thus yielding a tighter constraint on $H_0$. The precision of estimating $H_0$ can be improved from $2.1\%$ to $1.3\%$ with adding quasars as standard candles when the uncertainties of the time-delay distances are $5\%$.




\section{Conclusion and Discussions} \label{sec:con}


In this work, we develop a model-independent method to measure the Hubble constant with time-delay distances from strong lens systems combined with quasar and SNe Ia standard candles. 

We first generate mock data sets of SNe Ia, quasars, and  
strong lenses based on a fiducial cosmological model. 
Then we apply our GP regression technique on the mock SNe Ia and quasar data sets to simultaneously calibrate the mock quasar data set and reconstruct the Universe's expansion history. This reconstruction extends out to a redshift of 7.5, further than any other reconstruction to date. We also demonstrate how to test the reliability of our calibrated results, namely by calculating the normalized residuals of the $\log F_{\rm{X}}$ with respect to the mock SNe Ia data set. If the normalized residuals follow a Gaussian distribution, then the calibration results are reliable.

Since both the quasar and SNe Ia data sets are unanchored, we then anchor the reconstruction of expansion history from those data sets with the time-delay distances from the mock strong lens data set.
Previous model-independent reconstructions of the expansion history have only used SNe Ia. Since SNe Ia only extend to a redshift of 2.26, the model-independent reconstructions can only use the three strong lens systems with source redshifts less than 2.26. Using quasar as standard candles extends the redshift coverage over our model-independent reconstruction, and thus we can use nine strong lens systems in our analysis.  This yields a 1.3\% precision on $H_0$ in an optimistic case (5\% precision for strong lens distances) for a future strong lens data set and 2.4\% precision in a less optimistic case (10\% precision).

Fortunately, we will obtain more well-measured time-delay strong lens systems with the onset of cosmic surveys such as Roman, LSST, and Euclid. In addition to lensed quasars, strongly lensed transients, such as SN, are coming soon \citep{2022ChPhL..39k9801L}. With future time-delay distance measurements together with a larger and more precise quasar data set, one can obtain $H_0$ more precisely with the method described in this work and understand the $H_0$ tension better.

\section*{Acknowledgements}
 KL was supported by National Natural Science Foundation of China (NSFC) No. 12222302, 11973034 and y Funds for the Central Universities (Wuhan University 1302/600460081). X.Li was supported by NSFC No. 12003006, Hebei NSF No. A2020205002 and the fund of Hebei Normal University No. L2020B02. A.S. would like to acknowledge the support by National Research Foundation of Korea NRF2021M3F7A1082053, and the support of the Korea Institute for Advanced Study (KIAS) grant funded by the government of Korea. This work benefits from the high performance computing clusters at College of Physics, Hebei Normal University.

\acknowledgments

\bibliography{references}

\begin{thebibliography}{}
\expandafter\ifx\csname natexlab\endcsname\relax\def\natexlab#1{#1}\fi
\providecommand{\url}[1]{\href{#1}{#1}}
\providecommand{\dodoi}[1]{doi:~\href{http://doi.org/#1}{\nolinkurl{#1}}}
\providecommand{\doeprint}[1]{\href{http://ascl.net/#1}{\nolinkurl{http://ascl.net/#1}}}
\providecommand{\doarXiv}[1]{\href{https://arxiv.org/abs/#1}{\nolinkurl{https://arxiv.org/abs/#1}}}

\bibitem[{Ade {et~al.}(2014)}]{Planck:2013pxb}
Ade, P. A.~R., {et~al.} 2014, Astron. Astrophys., 571, A16,
  \dodoi{10.1051/0004-6361/201321591}

\bibitem[{Ade {et~al.}(2016)}]{Planck:2015fie}
---. 2016, Astron. Astrophys., 594, A13, \dodoi{10.1051/0004-6361/201525830}

\bibitem[{{Aghamousa} {et~al.}(2017){Aghamousa}, {Hamann}, \&
  {Shafieloo}}]{2017JCAP...09..031A}
{Aghamousa}, A., {Hamann}, J., \& {Shafieloo}, A. 2017, \jcap, 2017, 031,
  \dodoi{10.1088/1475-7516/2017/09/031}

\bibitem[{Aghanim {et~al.}(2020)}]{Planck:2018vyg}
Aghanim, N., {et~al.} 2020, Astron. Astrophys., 641, A6,
  \dodoi{10.1051/0004-6361/201833910}

\bibitem[{{Akeson} {et~al.}(2019){Akeson}, {Armus}, {Bachelet}, {Bailey},
  {Bartusek}, {Bellini}, {Benford}, {Bennett}, {Bhattacharya}, {Bohlin},
  {Boyer}, {Bozza}, {Bryden}, {Calchi Novati}, {Carpenter}, {Casertano},
  {Choi}, {Content}, {Dayal}, {Dressler}, {Dor{\'e}}, {Fall}, {Fan}, {Fang},
  {Filippenko}, {Finkelstein}, {Foley}, {Furlanetto}, {Kalirai}, {Gaudi},
  {Gilbert}, {Girard}, {Grady}, {Greene}, {Guhathakurta}, {Heinrich},
  {Hemmati}, {Hendel}, {Henderson}, {Henning}, {Hirata}, {Ho}, {Huff},
  {Hutter}, {Jansen}, {Jha}, {Johnson}, {Jones}, {Kasdin}, {Kelly}, {Kirshner},
  {Koekemoer}, {Kruk}, {Lewis}, {Macintosh}, {Madau}, {Malhotra}, {Mandel},
  {Massara}, {Masters}, {McEnery}, {McQuinn}, {Melchior}, {Melton},
  {Mennesson}, {Peeples}, {Penny}, {Perlmutter}, {Pisani}, {Plazas}, {Poleski},
  {Postman}, {Ranc}, {Rauscher}, {Rest}, {Roberge}, {Robertson}, {Rodney},
  {Rhoads}, {Rhodes}, {Ryan}, {Sahu}, {Sand}, {Scolnic}, {Seth}, {Shvartzvald},
  {Siellez}, {Smith}, {Spergel}, {Stassun}, {Street}, {Strolger}, {Szalay},
  {Trauger}, {Troxel}, {Turnbull}, {van der Marel}, {von der Linden}, {Wang},
  {Weinberg}, {Williams}, {Windhorst}, {Wollack}, {Wu}, {Yee}, \&
  {Zimmerman}}]{2019arXiv190205569A}
{Akeson}, R., {Armus}, L., {Bachelet}, E., {et~al.} 2019, arXiv e-prints,
  arXiv:1902.05569, \dodoi{10.48550/arXiv.1902.05569}

\bibitem[{{Alam} {et~al.}(2021){Alam}, {de Mattia}, {Tamone}, {{\'A}vila},
  {Peacock}, {Gonzalez-Perez}, {Smith}, {Raichoor}, {Ross}, {Bautista},
  {Burtin}, {Comparat}, {Dawson}, {du Mas des Bourboux}, {Escoffier},
  {Gil-Mar{\'\i}n}, {Habib}, {Heitmann}, {Hou}, {Mohammad}, {Mueller},
  {Neveux}, {Paviot}, {Percival}, {Rossi}, {Ruhlmann-Kleider}, {Tojeiro},
  {Vargas Maga{\~n}a}, {Zhao}, \& {Zhao}}]{2021MNRAS.504.4667A}
{Alam}, S., {de Mattia}, A., {Tamone}, A., {et~al.} 2021, \mnras, 504, 4667,
  \dodoi{10.1093/mnras/stab1150}

\bibitem[{{Collett} {et~al.}(2019){Collett}, {Montanari}, \&
  {R{\"a}s{\"a}nen}}]{2019PhRvL.123w1101C}
{Collett}, T., {Montanari}, F., \& {R{\"a}s{\"a}nen}, S. 2019, \prl, 123,
  231101, \dodoi{10.1103/PhysRevLett.123.231101}

\bibitem[{{Collett}(2015)}]{2015ApJ...811...20C}
{Collett}, T.~E. 2015, \apj, 811, 20, \dodoi{10.1088/0004-637X/811/1/20}

\bibitem[{Collett(2015)}]{Collett:2015roa}
Collett, T.~E. 2015, Astrophys. J., 811, 20, \dodoi{10.1088/0004-637X/811/1/20}

\bibitem[{Di~Valentino {et~al.}(2021)Di~Valentino, Mena, Pan, Visinelli, Yang,
  Melchiorri, Mota, Riess, \& Silk}]{DiValentino:2021izs}
Di~Valentino, E., Mena, O., Pan, S., {et~al.} 2021, Class. Quant. Grav., 38,
  153001, \dodoi{10.1088/1361-6382/ac086d}

\bibitem[{Du {et~al.}(2023)Du, Wei, You, Chen, Zhu, \& Liang}]{Du:2023zsz}
Du, S.-S., Wei, J.-J., You, Z.-Q., {et~al.} 2023, Mon. Not. Roy. Astron. Soc.,
  521, 4963, \dodoi{10.1093/mnras/stad696}

\bibitem[{Ertl {et~al.}(2023)Ertl, Schuldt, Suyu, Schmidt, Treu, Birrer,
  Shajib, \& Sluse}]{Ertl:2022rqx}
Ertl, S., Schuldt, S., Suyu, S.~H., {et~al.} 2023, Astron. Astrophys., 672, A2,
  \dodoi{10.1051/0004-6361/202244909}

\bibitem[{Foreman-Mackey {et~al.}(2013)Foreman-Mackey, Hogg, Lang, \&
  Goodman}]{foreman2013emcee}
Foreman-Mackey, D., Hogg, D.~W., Lang, D., \& Goodman, J. 2013, Publications of
  the Astronomical Society of the Pacific, 125, 306

\bibitem[{Holsclaw {et~al.}(2010{\natexlab{a}})Holsclaw, Alam, Sanso, Lee,
  Heitmann, Habib, \& Higdon}]{Holsclaw:2010nb}
Holsclaw, T., Alam, U., Sanso, B., {et~al.} 2010{\natexlab{a}}, Phys. Rev. D,
  82, 103502, \dodoi{10.1103/PhysRevD.82.103502}

\bibitem[{Holsclaw {et~al.}(2010{\natexlab{b}})Holsclaw, Alam, Sanso, Lee,
  Heitmann, Habib, \& Higdon}]{Holsclaw:2010sk}
---. 2010{\natexlab{b}}, Phys. Rev. Lett., 105, 241302,
  \dodoi{10.1103/PhysRevLett.105.241302}

\bibitem[{{Holsclaw} {et~al.}(2011){Holsclaw}, {Alam}, {Sans{\'o}}, {Lee},
  {Heitmann}, {Habib}, \& {Higdon}}]{2011PhRvD..84h3501H}
{Holsclaw}, T., {Alam}, U., {Sans{\'o}}, B., {et~al.} 2011, \prd, 84, 083501,
  \dodoi{10.1103/PhysRevD.84.083501}

\bibitem[{Hounsell {et~al.}(2023)Hounsell, Scolnic, Brout,
  {et~al.}}]{Hounsell:2023xds}
Hounsell, R., Scolnic, D., Brout, D., {et~al.} 2023.
\newblock \doarXiv{2307.02670}

\bibitem[{{Hwang} {et~al.}(2023){Hwang}, {L'Huillier}, {Keeley}, {Jee}, \&
  {Shafieloo}}]{2023JCAP...02..014H}
{Hwang}, S.-g., {L'Huillier}, B., {Keeley}, R.~E., {Jee}, M.~J., \&
  {Shafieloo}, A. 2023, \jcap, 2023, 014, \dodoi{10.1088/1475-7516/2023/02/014}

\bibitem[{{Joudaki} {et~al.}(2018){Joudaki}, {Kaplinghat}, {Keeley}, \&
  {Kirkby}}]{2018PhRvD..97l3501J}
{Joudaki}, S., {Kaplinghat}, M., {Keeley}, R., \& {Kirkby}, D. 2018, \prd, 97,
  123501, \dodoi{10.1103/PhysRevD.97.123501}

\bibitem[{Keeley {et~al.}(2021)Keeley, Shafieloo, Zhao, Vazquez, \&
  Koo}]{Keeley:2020aym}
Keeley, R.~E., Shafieloo, A., Zhao, G.-B., Vazquez, J.~A., \& Koo, H. 2021,
  Astron. J., 161, 151, \dodoi{10.3847/1538-3881/abdd2a}

\bibitem[{Khadka \& Ratra(2020)}]{Khadka:2020vlh}
Khadka, N., \& Ratra, B. 2020, Mon. Not. Roy. Astron. Soc., 497, 263,
  \dodoi{10.1093/mnras/staa1855}

\bibitem[{Khadka \& Ratra(2021)}]{Khadka:2020tlm}
---. 2021, Mon. Not. Roy. Astron. Soc., 502, 6140,
  \dodoi{10.1093/mnras/stab486}

\bibitem[{Li {et~al.}(2021)Li, Keeley, Shafieloo, Zheng, Cao, Biesiada, \&
  Zhu}]{Li:2021onq}
Li, X., Keeley, R.~E., Shafieloo, A., {et~al.} 2021, Mon. Not. Roy. Astron.
  Soc., 507, 919, \dodoi{10.1093/mnras/stab2154}

\bibitem[{{Liao} {et~al.}(2022){Liao}, {Biesiada}, \&
  {Zhu}}]{2022ChPhL..39k9801L}
{Liao}, K., {Biesiada}, M., \& {Zhu}, Z.-H. 2022, Chinese Physics Letters, 39,
  119801, \dodoi{10.1088/0256-307X/39/11/119801}

\bibitem[{Liao {et~al.}(2019)Liao, Shafieloo, Keeley, \& Linder}]{Liao:2019qoc}
Liao, K., Shafieloo, A., Keeley, R.~E., \& Linder, E.~V. 2019, Astrophys. J.
  Lett., 886, L23, \dodoi{10.3847/2041-8213/ab5308}

\bibitem[{Liao {et~al.}(2020)Liao, Shafieloo, Keeley, \& Linder}]{Liao:2020zko}
---. 2020, Astrophys. J. Lett., 895, L29, \dodoi{10.3847/2041-8213/ab8dbb}

\bibitem[{Lusso \& Risaliti(2017)}]{Lusso:2017hgz}
Lusso, E., \& Risaliti, G. 2017, Astron. Astrophys., 602, A79,
  \dodoi{10.1051/0004-6361/201630079}

\bibitem[{Lusso {et~al.}(2020)}]{Lusso:2020pdb}
Lusso, E., {et~al.} 2020, Astron. Astrophys., 642, A150,
  \dodoi{10.1051/0004-6361/202038899}

\bibitem[{{Mortlock} {et~al.}(2011){Mortlock}, {Warren}, {Venemans}, {Patel},
  {Hewett}, {McMahon}, {Simpson}, {Theuns}, {Gonz{\'a}les-Solares}, {Adamson},
  {Dye}, {Hambly}, {Hirst}, {Irwin}, {Kuiper}, {Lawrence}, \&
  {R{\"o}ttgering}}]{2011Natur.474..616M}
{Mortlock}, D.~J., {Warren}, S.~J., {Venemans}, B.~P., {et~al.} 2011, \nat,
  474, 616, \dodoi{10.1038/nature10159}

\bibitem[{{Oguri} \& {Marshall}(2010)}]{2010MNRAS.405.2579O}
{Oguri}, M., \& {Marshall}, P.~J. 2010, \mnras, 405, 2579,
  \dodoi{10.1111/j.1365-2966.2010.16639.x}

\bibitem[{Rasmussen \& Williams(2006)}]{Rasmussen:2006}
Rasmussen, C.~E., \& Williams, C. K.~I. 2006, The MIT Press

\bibitem[{Reid {et~al.}(2019)Reid, Pesce, \& Riess}]{Reid_2019}
Reid, M.~J., Pesce, D.~W., \& Riess, A.~G. 2019, The Astrophysical Journal
  Letters, 886, L27, \dodoi{10.3847/2041-8213/ab552d}

\bibitem[{Riess {et~al.}(2019)Riess, Casertano, Yuan, Macri, \&
  Scolnic}]{Riess_2019}
Riess, A.~G., Casertano, S., Yuan, W., Macri, L.~M., \& Scolnic, D. 2019, The
  Astrophysical Journal, 876, 85, \dodoi{10.3847/1538-4357/ab1422}

\bibitem[{Riess {et~al.}(2018)Riess, Casertano, Yuan, Macri, Anderson,
  MacKenty, Bowers, Clubb, Filippenko, Jones, \& Tucker}]{Riess_2018}
Riess, A.~G., Casertano, S., Yuan, W., {et~al.} 2018, The Astrophysical
  Journal, 855, 136, \dodoi{10.3847/1538-4357/aaadb7}

\bibitem[{Riess {et~al.}(2022)}]{Riess:2021jrx}
Riess, A.~G., {et~al.} 2022, Astrophys. J. Lett., 934, L7,
  \dodoi{10.3847/2041-8213/ac5c5b}

\bibitem[{Risaliti \& Lusso(2015)}]{Risaliti:2015zla}
Risaliti, G., \& Lusso, E. 2015, Astrophys. J., 815, 33,
  \dodoi{10.1088/0004-637X/815/1/33}

\bibitem[{Risaliti \& Lusso(2019)}]{Risaliti:2018reu}
---. 2019, Nature Astron., 3, 272, \dodoi{10.1038/s41550-018-0657-z}

\bibitem[{{Schlegel} {et~al.}(2009){Schlegel}, {White}, \&
  {Eisenstein}}]{2009astro2010S.314S}
{Schlegel}, D., {White}, M., \& {Eisenstein}, D. 2009, in astro2010: The
  Astronomy and Astrophysics Decadal Survey, Vol. 2010, 314.
\newblock \doarXiv{0902.4680}

\bibitem[{Schmidt {et~al.}(2023)}]{DES:2022dvw}
Schmidt, T., {et~al.} 2023, Mon. Not. Roy. Astron. Soc., 518, 1260,
  \dodoi{10.1093/mnras/stac2235}

\bibitem[{Scolnic {et~al.}(2022)}]{Scolnic:2021amr}
Scolnic, D., {et~al.} 2022, Astrophys. J., 938, 113,
  \dodoi{10.3847/1538-4357/ac8b7a}

\bibitem[{Shafieloo {et~al.}(2012)Shafieloo, Kim, \&
  Linder}]{Shafieloo2012Gaussian}
Shafieloo, A., Kim, A.~G., \& Linder, E.~V. 2012, Physical Review D, 85, 123530

\bibitem[{{Shafieloo} {et~al.}(2013){Shafieloo}, {Kim}, \&
  {Linder}}]{2013PhRvD..87b3520S}
{Shafieloo}, A., {Kim}, A.~G., \& {Linder}, E.~V. 2013, \prd, 87, 023520,
  \dodoi{10.1103/PhysRevD.87.023520}

\bibitem[{{Spergel} {et~al.}(2015){Spergel}, {Gehrels}, {Baltay}, {Bennett},
  {Breckinridge}, {Donahue}, {Dressler}, {Gaudi}, {Greene}, {Guyon}, {Hirata},
  {Kalirai}, {Kasdin}, {Macintosh}, {Moos}, {Perlmutter}, {Postman},
  {Rauscher}, {Rhodes}, {Wang}, {Weinberg}, {Benford}, {Hudson}, {Jeong},
  {Mellier}, {Traub}, {Yamada}, {Capak}, {Colbert}, {Masters}, {Penny},
  {Savransky}, {Stern}, {Zimmerman}, {Barry}, {Bartusek}, {Carpenter}, {Cheng},
  {Content}, {Dekens}, {Demers}, {Grady}, {Jackson}, {Kuan}, {Kruk}, {Melton},
  {Nemati}, {Parvin}, {Poberezhskiy}, {Peddie}, {Ruffa}, {Wallace}, {Whipple},
  {Wollack}, \& {Zhao}}]{2015arXiv150303757S}
{Spergel}, D., {Gehrels}, N., {Baltay}, C., {et~al.} 2015, arXiv e-prints,
  arXiv:1503.03757, \dodoi{10.48550/arXiv.1503.03757}

\bibitem[{Treu {et~al.}(2022)Treu, Suyu, \& Marshall}]{Treu:2022aqp}
Treu, T., Suyu, S.~H., \& Marshall, P.~J. 2022, Astron. Astrophys. Rev., 30, 8,
  \dodoi{10.1007/s00159-022-00145-y}

\end{thebibliography}
\end{document}